\documentclass{article}

\usepackage{PRIMEarxiv}

\usepackage[utf8]{inputenc} 
\usepackage[T1]{fontenc}    
\usepackage{hyperref}       
\usepackage{url}            
\usepackage{booktabs}       
\usepackage{amsfonts}       
\usepackage{nicefrac}       
\usepackage{microtype}      
\usepackage{lipsum}
\usepackage{graphicx}
\graphicspath{{images/}}     

\title{The Effects of Just-in-time Delivery on Social Engagement: A Cluster Analysis
}

\author{
M. Ramírez \\
  Universidad Tecnológica de la Mixteca \\
  Huajuapan de León, Oaxaca \\
  México\\
  \texttt{merg@mixteco.utm.mx}
   \And
  R. Ruíz, N. Klarer \\
  Datyra Inc. \\
   San Diego, California \\
  U.S.A.\\
  \texttt{\{raziel, n\}@datyra.com}   
}

\begin{document}
\maketitle

\begin{abstract}

Fooji Inc. is a social media engagement platform that has created a proprietary "Just-in-time" delivery network to provide prizes to social media marketing campaign participants in real-time. In this paper, we prove the efficacy of the "Just-in-time" delivery network through a cluster analysis that extracts and presents the underlying drivers of campaign engagement.

We utilize a machine learning methodology with a principal component analysis to organize Fooji campaigns across these principal components. The arrangement of data across the principal component space allows us to expose underlying trends using a $K$-means clustering technique.  The most important of these trends is the demonstration of how the "Just-in-time" delivery network improves social media engagement.
\end{abstract}



\section{Introduction}

Fooji is a fan engagement platform that provides the world's largest brands the capability to on-deman prizes to participants in social media campaigns. This paper describes the analysis of the Fooji historical campaign performance data set (FCD). This data set contains historical campaign data, provided by Fooji Inc., analyzed with the goal of determining whether "Just-in-time" deliver increases social engagement in marketing campaigns. 

As a first step, we preprocessed the data to clean and impute missing values. A PostgreSQL database server is used to store the original data set. Using a SQLalchemy connector the data was extracted; numerical libraries like Pandas and Numpy were used to preprocess the data and to obtain clean data (Molin 2019). 

In order to maximize the number of observations in the data set, a $k-nn$ imputation algorithm was applied to impute missing values.

The FCD was subsequently analyzed using a Principal Component Analysis (PCA) to reduce the dimensionality of the data and determine variable importance \cite{haykin2009neural}.  Subsequentally, the campaigns were clustered across their principal components and the attributes of the clusters were reviewed to determine whether there were observable differences between the clusters.



\section{Methodology}

\begin{figure}[hbt!]
\centering
\includegraphics[width=0.60\linewidth]{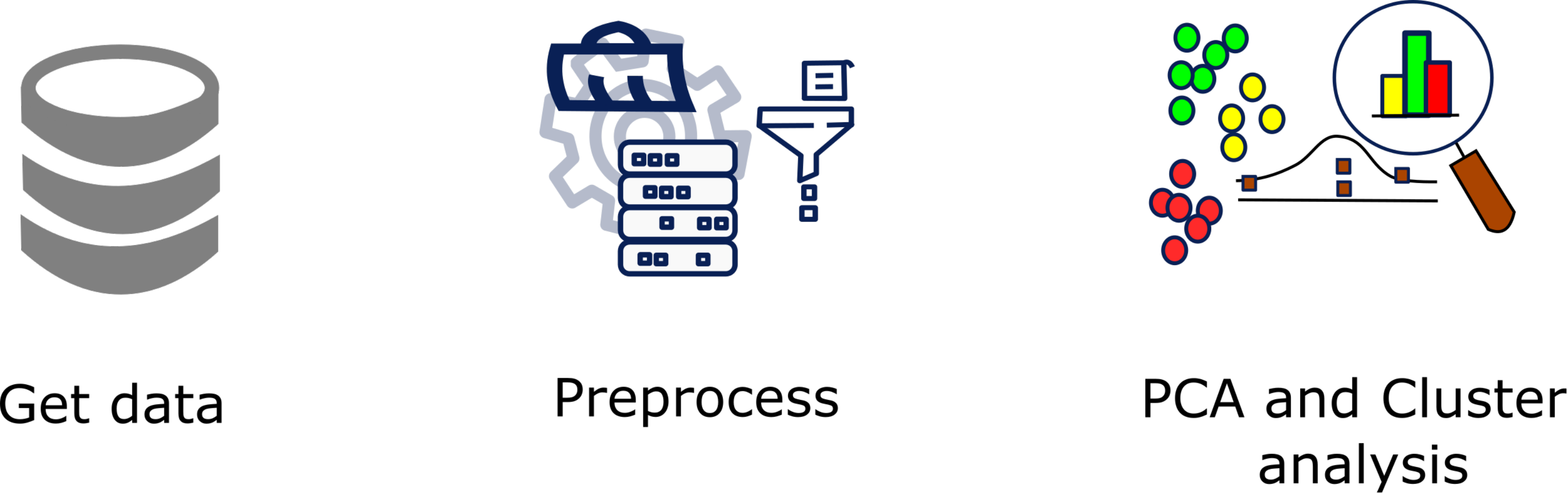}
\caption{Methodology to implement ML algorithms.}
\label{fig_MLpipeline}
\end{figure}

The general steps in our experiment are shown in Figure \ref{fig_MLpipeline}. As a first step, the data set  was obtained from several sources in Fooji databases. The data shared from Fooji Inc. was exported to a Postgres database running on an AWS service. All data was processed on an EC2 instance. 

After the preprocessing is done, the clean data set is used to perform our PCA and Cluster analysis. We used measured the performance of our PCA analysis in terms of the percentage of variance captured. Then, we measured the performance of our cluster analysis using the Elbow Method \cite{syakur2018integration}.

\subsection{Preprocessing}

\begin{figure}[hbt]
\centering
\includegraphics[width=0.650\linewidth]{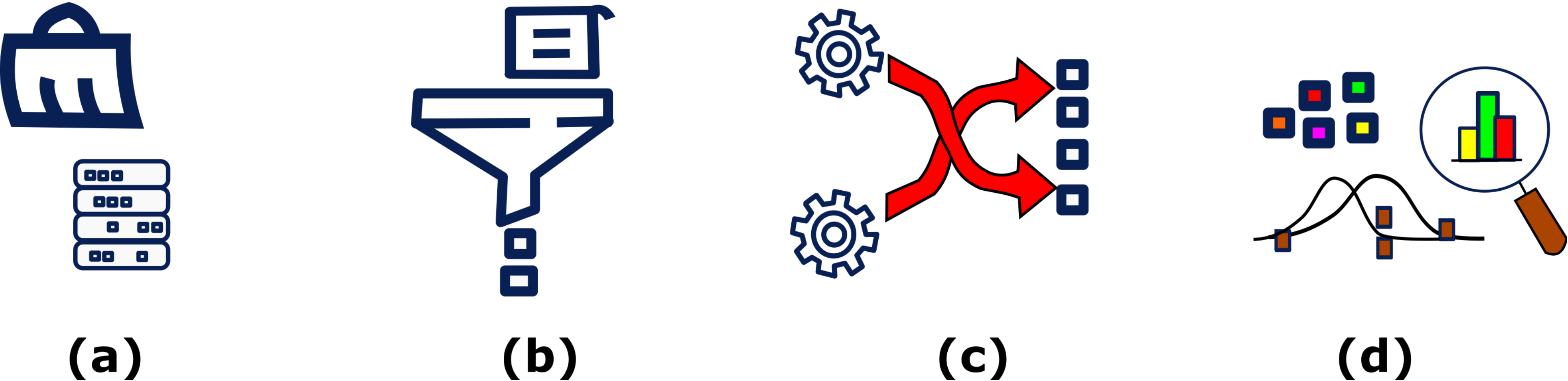}
\caption{Preprocessing pipeline.
(a) Cleansing of the data, (b) Filtering for useful values, (c) Transformations may imply principal component analysis, missing data analysis, feature selection, among others, (d) Validation can be done via graphical analysis of the features in order to be sure that this phase generates coherent and correct results.}
\label{fig_preprocessing}
\end{figure}

Preprocessing the data is a critical part of all the machine learning processes. High-quality utilization of numerical tools to clean the data may help to improve the performance of the ML algorithms. The output of this phase generates the data to be used in the analysis and later to test the analysis using our performance metrics.  Preprocessing of the data includes several phases (see Figure \ref{fig_preprocessing}). As a first phase while preprocessing, a cleaning of the data was performed. Several tables were created to analyze and summarize the elements to be used in the next phases.

\subsection{Cleansing and filtering}
As a first step for preprocessing a cleansing and filtering of the records was done. The data set had three distinct dependent variables used to measure campaign outcomes: \textit{Messages}, \textit{Participations} and \textit{Engagements}. The list of independent variables is provided in Table \ref{table_independent_variables}.
 It’s important to mention that we computed distinct record count for every dependent variable because some independent variables were missing values. For \textit{messages} dependent variable, the total number of records in the data set is 907, for \textit{participations} 654 and for \textit{engagements}  612.

\begin{table}[hbt!]
\centering
\begin{tabular}{llll}
\hline
    Description &  Independent variable \\
\hline
    Fulfillment method category & \texttt{fm\_categories} \\
    Give away method category & \texttt{giveaways\_categories} \\ 
    Campaign flight total days & \texttt{cf\_total\_days} \\ 
    Allocated deliveries total & \texttt{ad\_total} \\ 
    Price type category & \texttt{pt\_categories} \\ 
    Social channel category &  \texttt{sch\_categories}     \\ 
\hline
\end{tabular}
\caption{Independent variables detected as the most important.
\label{table_independent_variables}
}
\end{table}

\subsection{Transformations}
Most machine learning (ML) algorithms need to have a dense vector of inputs. ML algorithms have a better performance if they are trained with a large number of records. As discussed above, several independent variables had missing values, and these records can’t be used by the ML algorithms; in order to keep as many records as possible and to be able to use them, a $k-nn$ imputation method 
was used to predict the missing values \cite{geron2022hands}. When an element of the record is missing, the $k-nn$ imputter searches for the $k$ elements that are most similar to the record with missing values, the missing data is approximated by the average of the fields that are the most similar \cite{hastie2009elements}.

\begin{table}[hbt!]
\centering
\begin{tabular}{llll}
\hline
\# & Column   & Non-null count  \\
\hline
 0  & campaign             & 907 non-null    \\
 1  & cf\_total\_days        & 907 non-null     \\
 2  & cf\_total\_seconds     & 907 non-null     \\
 3  & cf\_type              & 907 non-null     \\
 4  & giveaways\_categories & 646 non-null    \\
 5  & givaways\_totals      & 646 non-null   \\
 6  & pt\_categories        & 907 non-null     \\
 7  & pt\_totals            & 907 non-null   \\
 8  & fm\_categories        & 729 non-null    \\
 9  & fm\_total             & 729 non-null   \\
 10 & sch\_categories       & 907 non-null     \\
 11 & sch\_total            & 907 non-null   \\
 12 & ad\_categories        & 591 non-null    \\
 13 & ad\_total             & 591 non-null   \\
 14 & schm\_total           & 907 non-null   \\
\hline
\end{tabular}
\caption{Columns in the FCD to be used by the regression ML algorithm. 
\label{table_data_set}
}
\end{table}

Table \ref{table_data_set} contains the name of the campaign at column 0, independent variables are listed from  columns 1 to 13, column 14 contains the messages dependent variable, they can be used as input and output, respectively to train models. Columns 4, 5, 8, 9, 12 and 13 contain missing values. Categorical variables are named by using the word \texttt{categories} or \texttt{type}, numerical variables contain the word \texttt{total}. 

\subsection{Outliers and statistical metrics }

\begin{figure}[hbt!]
\centering
\includegraphics[width=0.750\linewidth]{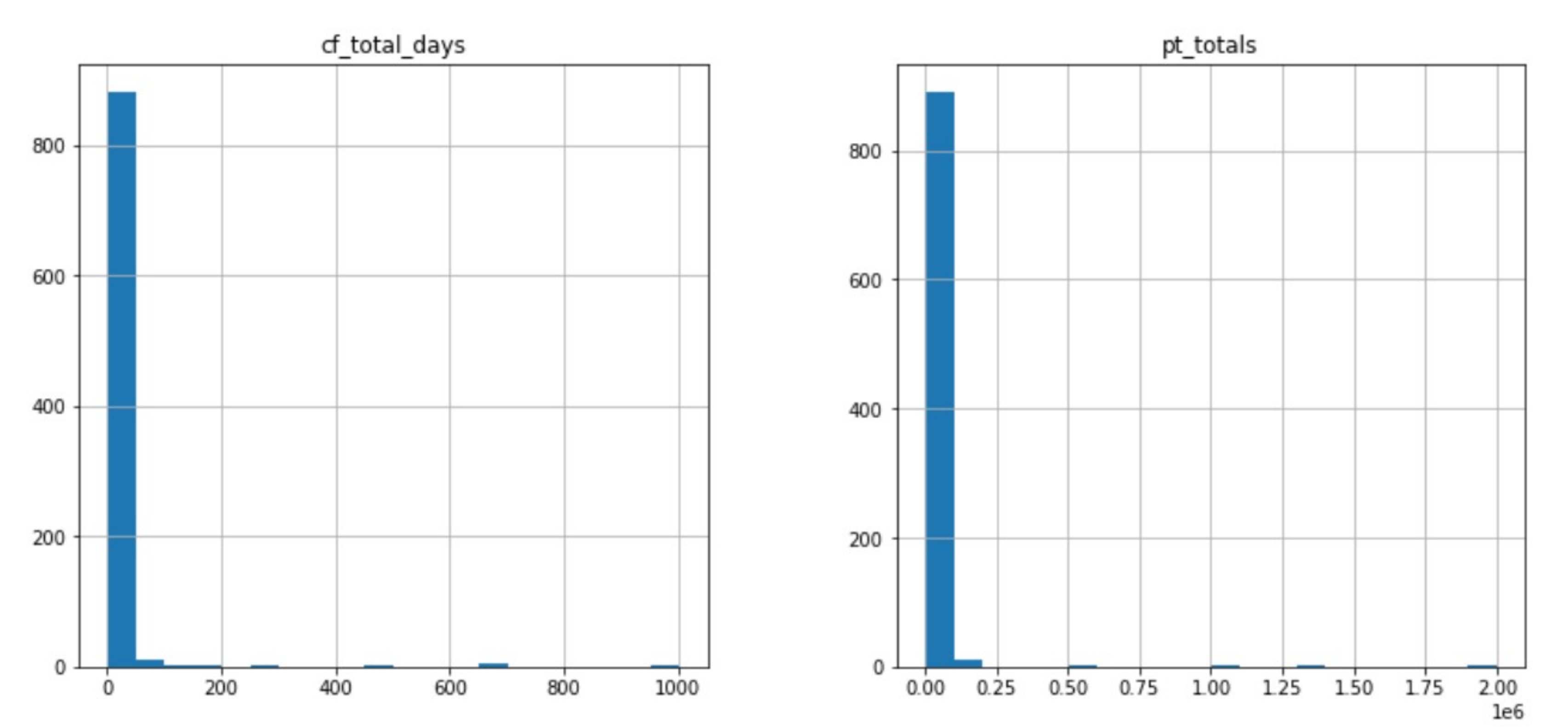}
\caption{Histograms of the distribution of data for the \texttt{cf\_total\_seconds} and \texttt{pt\_totals} independent variables. The other independent numerical variables have a similar behavior.
\label{fig_histograms_distributions}
}
\end{figure}

Outliers are data points that are distant from the rest\footnote{\url{https://www.neuraldesigner.com/blog/3\_methods\_to\_deal\_with\_outliers}}. These values may represent errors in measurements, bad data collection or simply high variability in the observations that are far from the principal population. ML algorithms are susceptible to outliers and can mislead the training process generating longer training times, and less accurate models. To analyze the outliers, a histogram was drawn to show the distribution of the values for numerical variables.  Figure \ref{fig_histograms_distributions} shows that two of the seven numerical variables are highly skewed\footnote{ Skewness is a measure of the lack of symmetry. A data set is symmetric if it looks the same to the left and right of the center point.
} to the left. This means that the distributions of the variables are concentrated in the lowest values at the left of the range of the data. 

The kurtosis is a measure of whether the data are heavy-tailed or light-tailed relative to a normal distribution. That is, data sets with high kurtosis tend to have heavy tails, or outliers \footnote{\url{https://www.itl.nist.gov/div898/handbook/eda/section3/eda35b.htm}}. In FCD, all independent variables show high kurtosis values, it implies that there’s a high presence of outliers. Table \ref{table_Kurt_Skew} shows the values for the numerical independent variables.

\begin{table}[hbt!]
\centering
\begin{tabular}{lllll}
 \  & \multicolumn{2}{c}{ \textbf{Original data}} & \multicolumn{2}{c}{ \textbf{After removing outliers}} \\
 \textbf{Independent variables}                        &  \textbf{Kurtosis} &  \textbf{Skewness} &  \textbf{Kurtosis}      &  \textbf{Skewness}    \\
cf\_total\_seconds                                                                & 130.29          & 8.97            & 44.65                & 5.21                 \\
cf\_total\_days                                                                   & 108.01          & 9.98            & 101.01               & 9.76                 \\
givaways\_totals                                                                  & 168.89          & 12.31           & 212.32               & 14.16                \\
pt\_totals                                                                        & 238.28          & 14.61           & 212.39               & 14.15                \\
fm\_total                                                                         & 190.58          & 13.07           & 212.32               & 14.16                \\
sch\_total                                                                        & 169.63          & 11.56           & 1.14                 & 1.22                 \\
ad\_total                                                                         & 535.18          & 22.62           & 314.46               & 17.23               
\end{tabular}
\caption{Kurtosis and skewness values for the numerical independent variables.
\label{table_Kurt_Skew}
}
\end{table}

\begin{figure}[hbt!]
\centering
\includegraphics[width=0.75\linewidth]{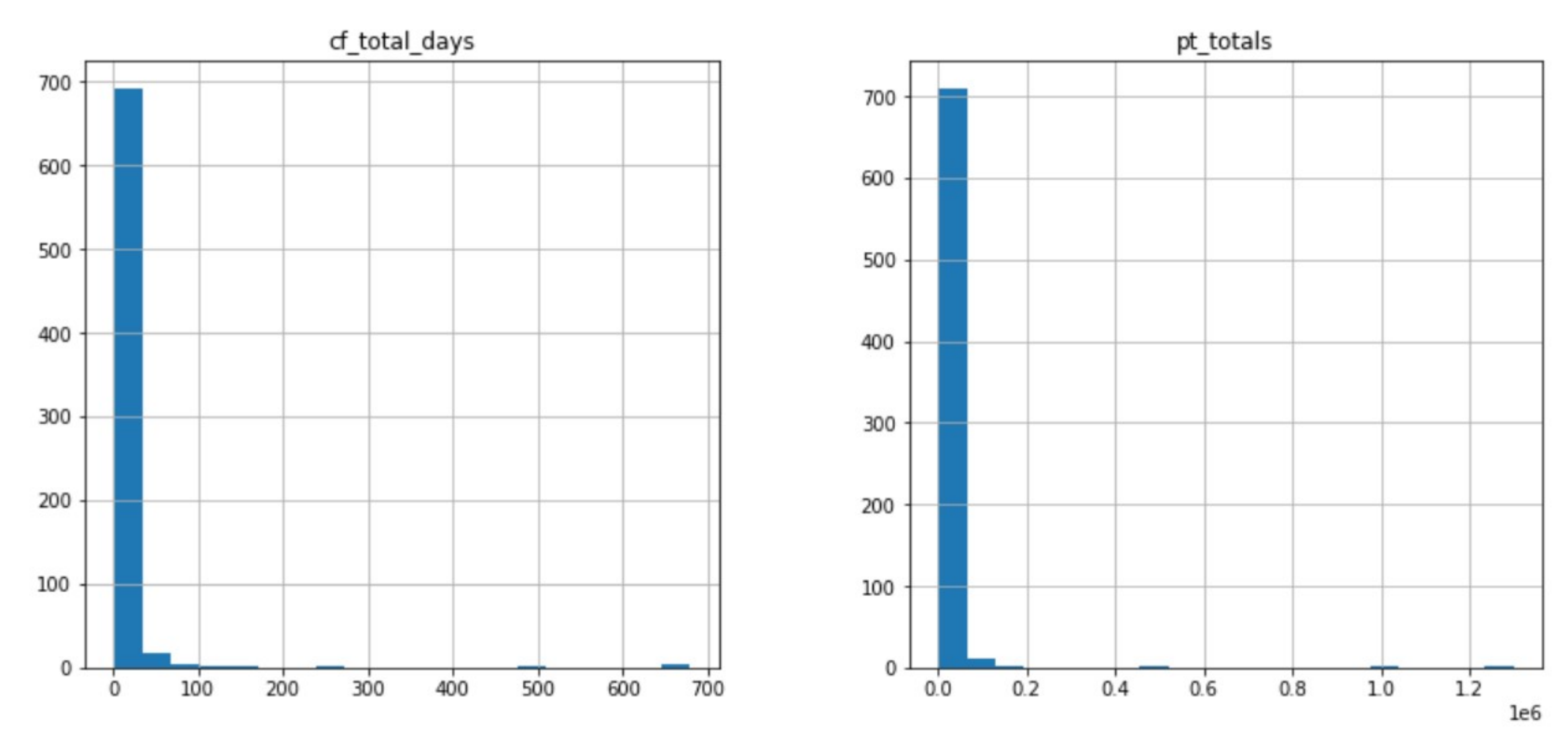}
\caption{ Histograms for  \texttt{cf\_total\_days} and \texttt{pt\_totals} show a decrease in values for frequency in the left parts of the data, also the higher values have been removed.
\label{fig_histograms_distributions_removed}
}
\end{figure}

Outlier deletion was done by removing the lower and higher 10\% of the data in the data set in every numerical variable. 
The result of removing outliers from  the \texttt{cf\_total\_days}  and \texttt{pt\_totals} variables in Figure  \ref{fig_histograms_distributions}  can be seen in Figure  \ref{fig_histograms_distributions_removed}.
In this image, the range of the values has decreased at the highest values, also the frequency in the lower values has decreased. This has decreased the values for kurtosis and skewness as shown in the right part of the Table \ref{table_Kurt_Skew}.

\subsection{Correlation in the variables}

After removing the outliers, a correlation analysis was performed to visualize the relationships among the  independent variables and the dependent variables. Before analyzing these relationships, categorical variables are transformed to numerical values. Additionally, all numerical variables are normalized in order to have a common scale in the data. If data is not normalized the distorting differences in the range of values may affect to the ML algorithms.

\subsection{Label encoders}

Commonly,  data sets contain numerical and categorical variables. Categorical variables can be in the form of words or numbers; to make the data understandable to ML algorithms these variables need to be converted into numbers. As the data set presents several categorical variables, a label encoding of these variables was important to translate all possible categorical values into numerical values \cite{geron2022hands}. 


Large differences between the ranges of numerical variables may cause variables with larger ranges to dominate over those with small ranges leading to biased results. This is the reason to standardize all numerical variables by subtracting the mean and dividing by the standard deviation for each value of each variable \cite{geron2022hands}. Once the standardization is done, all the variables will be transformed to the same scale.

\clearpage


\section{Primary Analysis of Data Using Machine Learning Techniques}

In this section, we perform the analysis of the data to determine whether there are meaningful observable differences between clusters and whether we can extract underlying trends from those clusters. We conduct this analysis via Principal Component Analysis (PCA) and a subsequent Clustering Analysis (CA) on the data across the Principal Components.

\subsection{Principal Component Analysis}
Principal Component Analysis (PCA) is a statistical unsupervised technique used to explore the interrelations among a set of variables in order to identify its underlying structure. $PCA$ summarizes the original data set with a smaller number of representative variables that collectively explain most of the variability in the original set \cite{jolliffe2016principal}. This method is used to reduce the dimensionality of the FCD and increase its interpretability using a smaller number of representative variables that collectively explain most of the variability in the original set. Once the data is reduced, a plot is introduced to evaluate the relationships in the data \cite{gareth2013introduction}.

\begin{figure}[hbt!]
\centering
\includegraphics[width=0.76\linewidth]{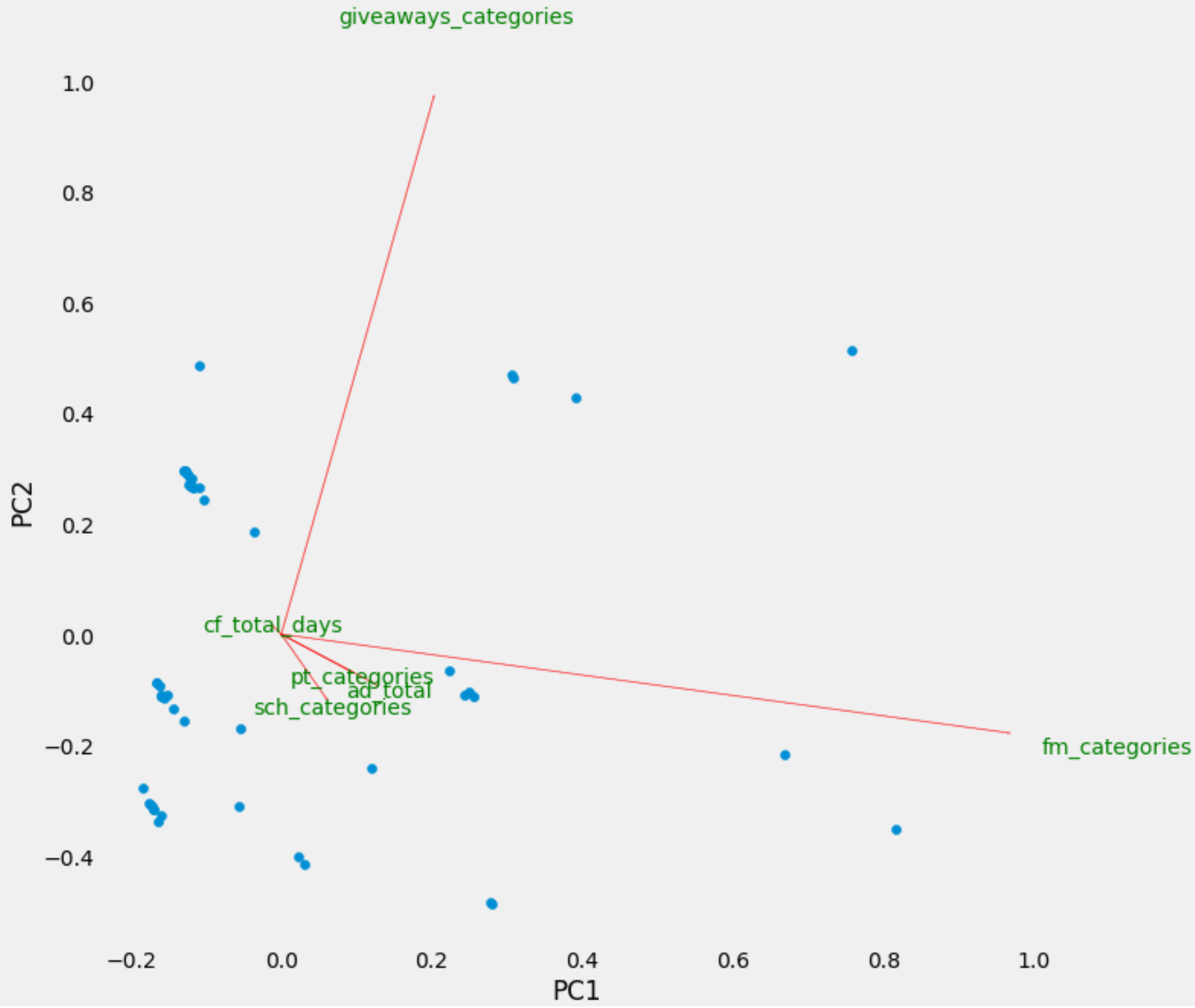}
\caption{ Plot of the data set using the first two components and the variables included in figure \ref{fig_correlations_categorical_vars}.
\label{fig_MagnitudePCA} }
\end{figure}

\begin{figure}[hbt!]
\centering
\includegraphics[width=0.60\linewidth]{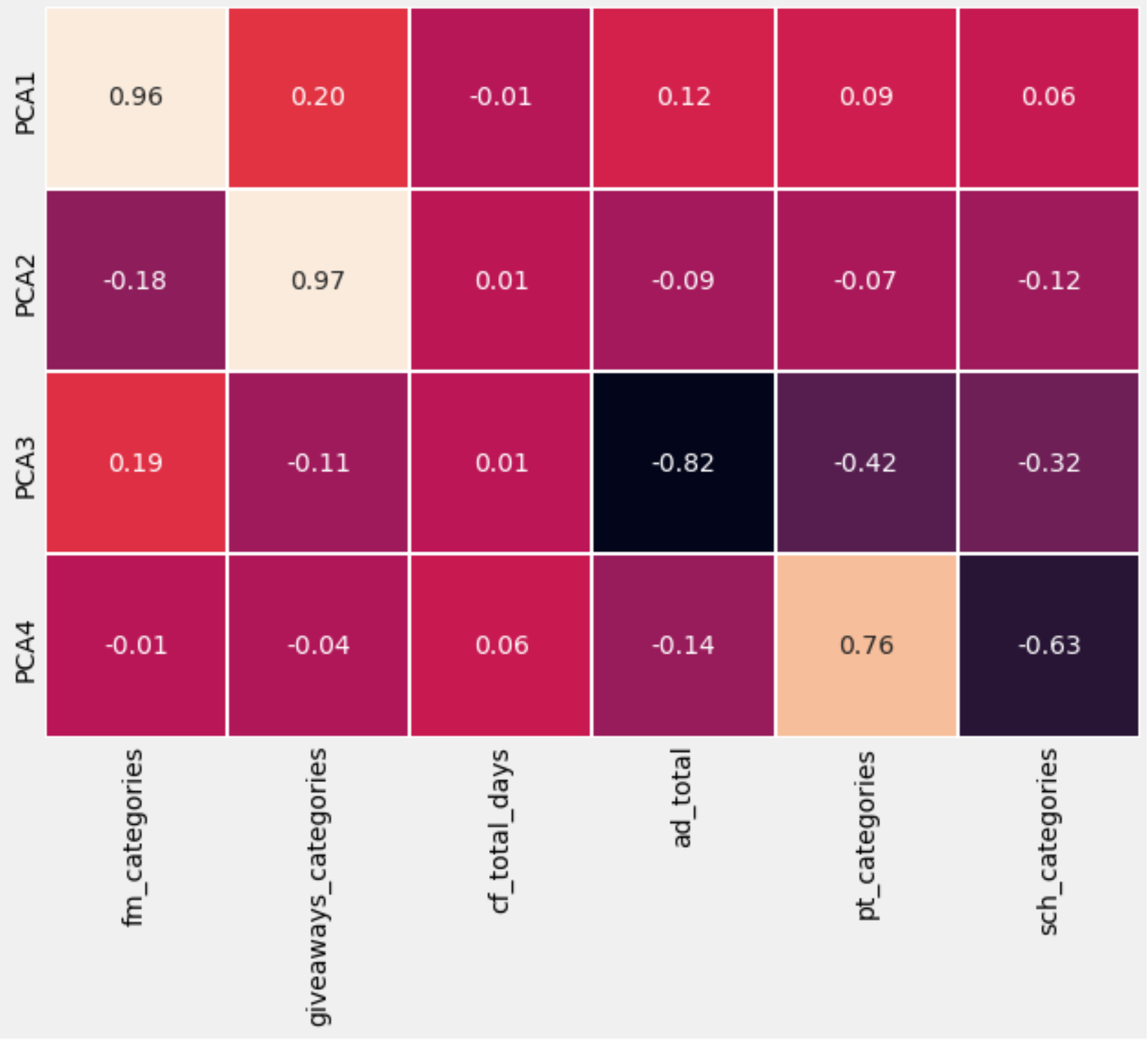}
\caption{ Correlation matrix of independent variables vs the total number of messages (\texttt{schm\_total}).  \label{fig_correlations_categorical_vars}
}
\end{figure}

We introduce a chart showing the first two principal components ($PCA_1$ and $PCA_2$) to visualize the importance of the independent variables in the overall $PCA$ shown in Figure \ref{fig_MagnitudePCA};  $PCA_1$ and $PCA_2$ are displayed on the  $X$ and $Y$ axis respectively. The blue points represent every observation, these observations are drawn using only the first $2$ components on the test data set.  
The magnitude of the red lines show the importance of every independent variable 
shown in Figure \ref{fig_MagnitudePCA}  and listed on the Table \ref{table_independent_variables}.
The magnitude of the independent variables is consistent with the values in the correlation matrix shown in figure \ref{fig_correlations_categorical_vars}; the two most important independent variables are Fulfillment Method and Giveaway Method respectively. The first component ($PC_1$) has a high correlation with \texttt{fm\_categories}. The second component has a high  relationship with \texttt{giveaways\_categories}. Remember, these variables are categorical and the direction of the correlation indicates a categorical preference. We will discuss the meaning of these categorical preferences in our analysis of the results.

\begin{figure}[hbt!]
\centering
\includegraphics[width=0.850\linewidth]{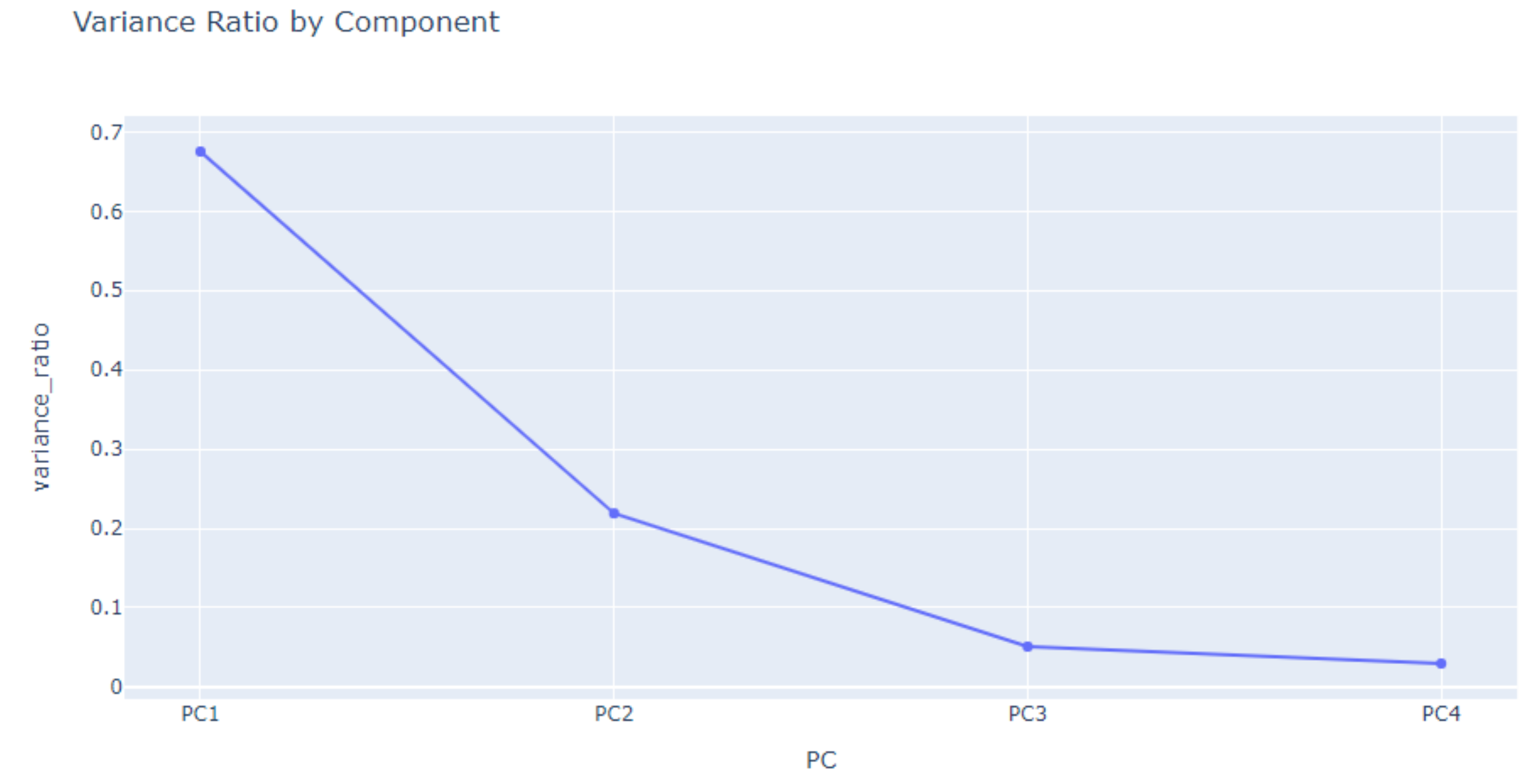}
\caption{ Variance ratio importance using PCA analysis.
\label{fig_variance_PCA}
}
\end{figure}

PCA algorithm distributes the variance of the data across its principal components. Figure \ref{fig_variance_PCA} shows the distribution of the importance in the first $4$ principal components. You can see that the first two Principal Components (with high correlations to Fulfillment Method and Giveaway Category) comprise nearly $90$ percent of the variance of the data. The $PCA$ algorithm attempts to minimize the total number of components required to summarize the data. In this case, we reach a total sum of the variance of $95\%$  using four Principal Components.

\subsection{Clustering analysis}

Cluster analysis (CA) is a technique to group similar observations into a number of meaningful clusters based on the observed values of the independent variables of the cluster. A cluster is meaningful if its elements are similar to one another and different from elements in other clusters.
CA is a quantitative form of classification \cite{giordani2020introduction}.  The classification procedures used in cluster analysis are based on either density of the population or the distance between members. These methods can serve to generate a basis for the classification of large numbers of dissimilar variables \cite{theodoridis2006pattern}.

\begin{figure}[hbt!]
\centering
\includegraphics[width=0.450\linewidth]{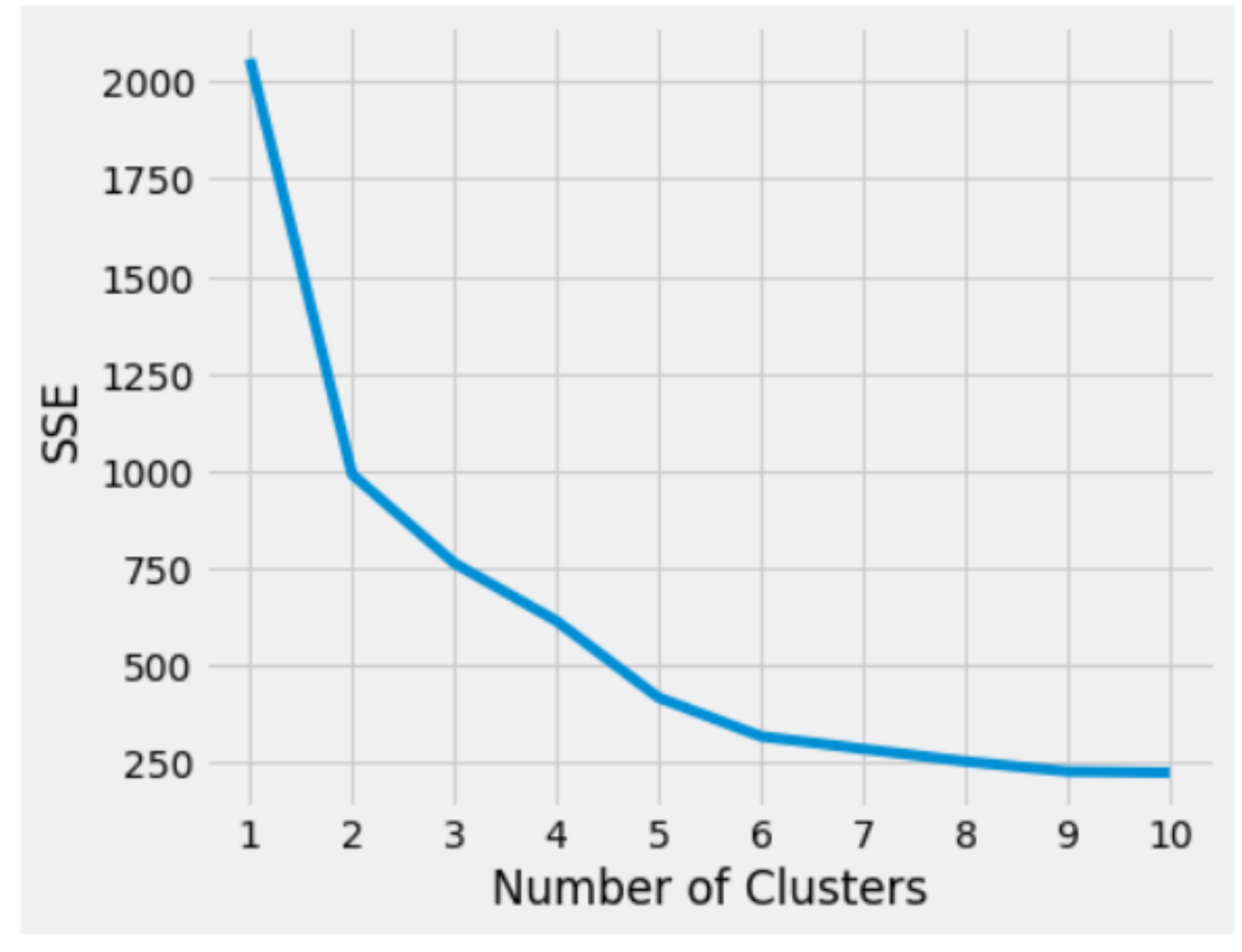}
\caption{ The elbow method obtained a total of $3$ clusters. \label{fig_elbow_method}
}
\end{figure}

$K$-means clustering is an unsupervised learning algorithm that groups data based on each point's euclidean distance to a central point called the centroid. The centroids are defined by the means of all points that are in the same cluster. The algorithm first chooses random points as centroids and then iterates adjusting them until full convergence \cite{theodoridis2006pattern,haykin2009neural}.  A fundamental step  in clustering is to determine the optimal number of clusters into which the data may be clustered. The Elbow Method is one of the most popular methods to determine this optimal value of $k$ by analyzing the inertia, this is the sum of squared distances of samples to their closest cluster center \cite{syakur2018integration}. To determine the optimal number of clusters, the value of $k$ at the ``elbow'' is selected as the point after which the inertia starts decreasing in a linear fashion. Thus for the given data, the conclusion is that the optimal number of clusters for the data is $3$, as shown in Figure \ref{fig_elbow_method}.

\begin{figure}[hbt!]
\centering
\includegraphics[width=0.985\linewidth]{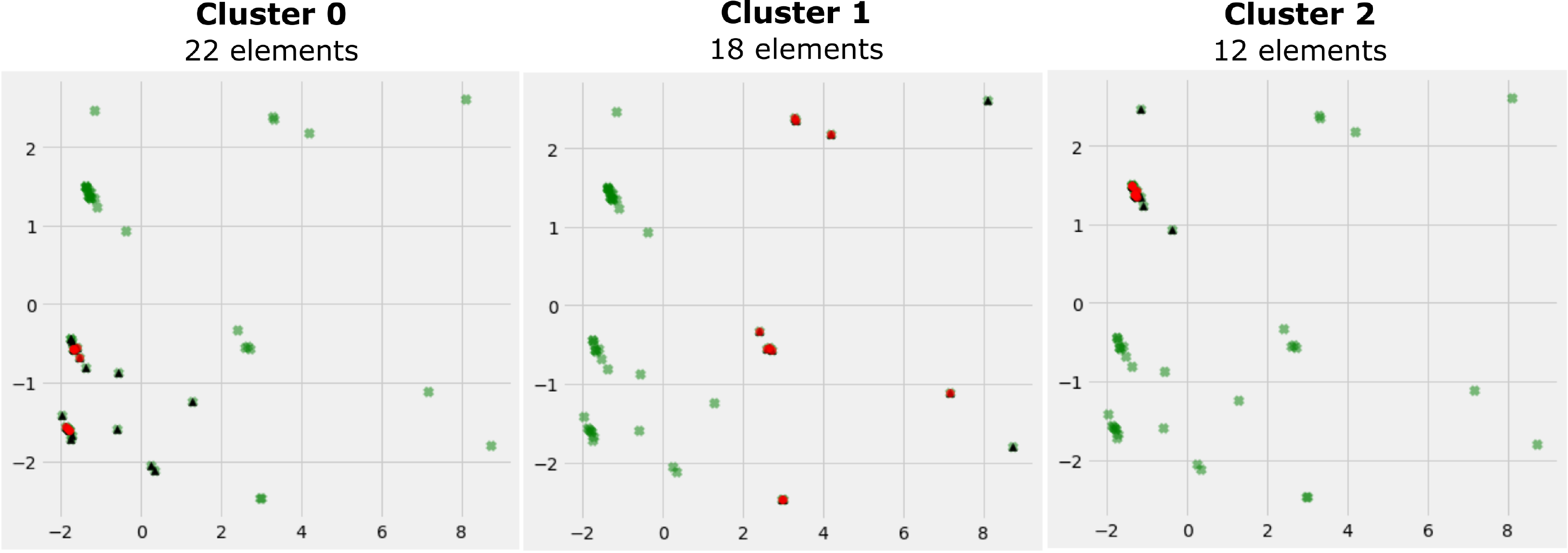}
\caption{ This is a visualization of the $3$ optimal clusters using the elbow method. The black points correspond to the labeled cluster, red points are the $10$ closest points to the centroid. Green values represent the rest of the population. The points are plotted using the first two principal components. The analysis clearly separates Cluster $1$ at a $PC_1$ value greater than $2$. This will be meaningful for our discussion of the results.
\label{fig_cluster_analysis}
}
\end{figure}

Figure \ref{fig_cluster_analysis} shows in cluster $1$ a group of outliers on the right side of the scatters plot, as it can be seen this data is distributed on the first principal component; this cluster is consistent with the histograms shown in Figure \ref{fig_histograms_distributions_removed}. Clusters $0$ and $2$ are distributed on the second principal component in two regions.

\subsection{Interpretation of Clusters}

\begin{figure}[hbt!]
\centering
\includegraphics[width=0.97\linewidth]{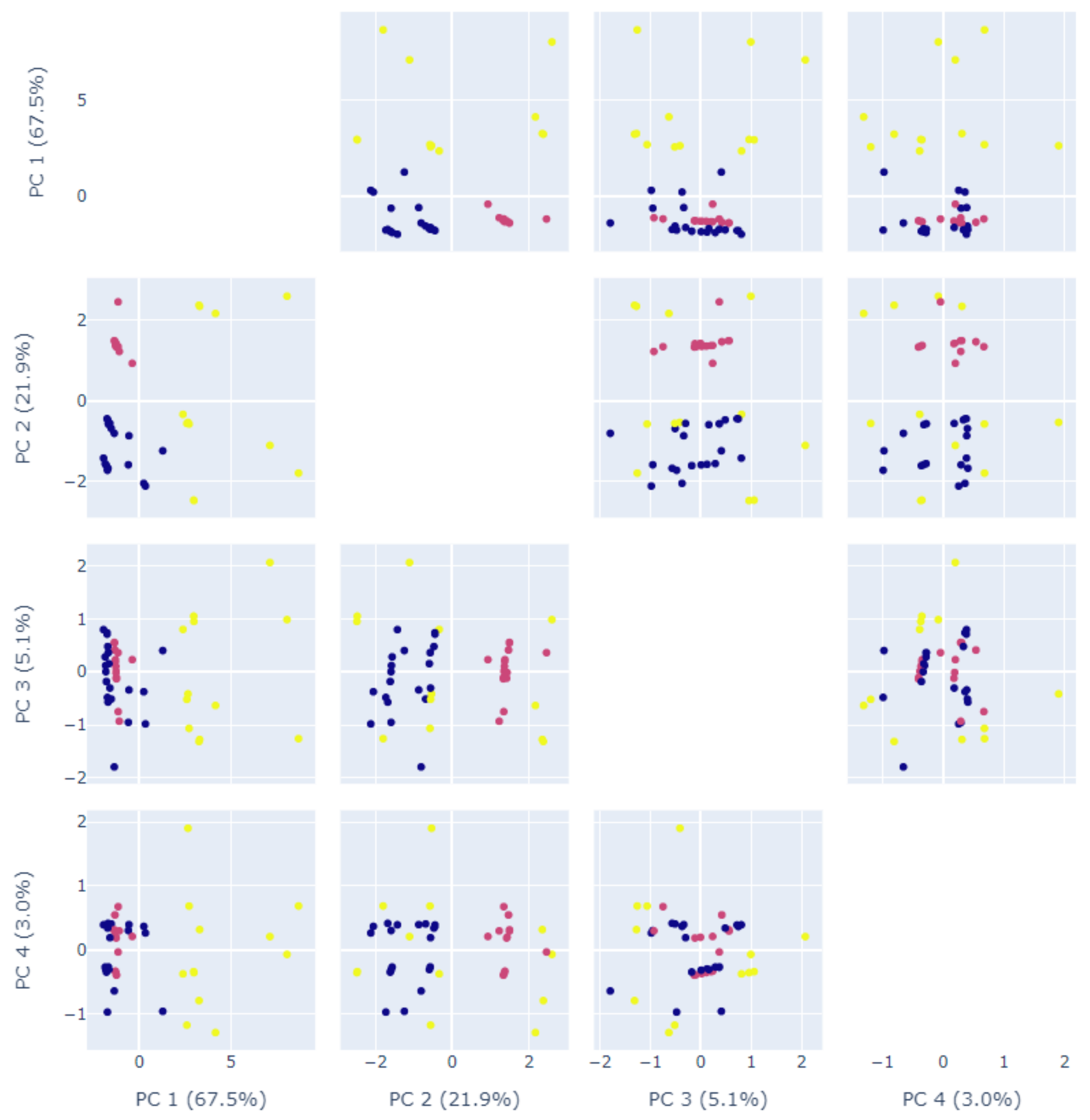}
\caption{ This pairplot shows the obtained clusters in three different colors (navy for cluster $0$, magenta for cluster $1$, and yellow for cluster $2$) from the FCD using PCA. Multidimensional entries for every campaign are mapped to the principal components shown on the charts. Every graph shows the data projected onto a plane of the principal components $PCA_i$ and $PCA_j$,$i \neq j$. 
\label{fig_clustering_result}
}
\end{figure}

Figure \ref{fig_clustering_result} shows how the observations are clustered in different groups or clusters. 
The first row represents the use of the first principal component, it shows a clear division of the regions or clusters of the data if this component is used.
By watching the chart on the second row and first column, $PCA_1$ divides the cluster $1$ (in yellow), from the clusters $0$ and $2$ (in navy and magenta, respectively); the $PCA_2$, in combination with the $PCA_1$ on the left region  can help to divide the data into the other two clusters.
Graphically, it can be seen  that charts in the rows and columns associated with the first principal component are more clearly separated, than when other combinations are used.
Even though the cluster centers are initialized in a random way the data is grouped in a coherent pattern.

Cluster $0$, as shown in Figure \ref{fig_cluster_analysis} contains data that is far from  the rest of the clusters. The dominant variable in $PC_1$ is the Fulfillment Method with the calculation being an admixture of the Fulfillment Method and other variables.

Clusters $1$ and $2$ have the same value for \texttt{fm\_categories}, but different values for  \texttt{giveaways\_categories}. Figure \ref{fig_correlations_categorical_vars} shows the $PCA_1$ and $PCA_2$ contain a high relationship to the variables \texttt{fm\_categories} and \texttt{giveaways\_categories}, respectively. Other variables have a lower impact on the principal components used to delineate the clusters. The impact of the Fulfillment Method and Giveaway Category on the determination of the clusters will be discussed in our conclusion.




\section{Analysis of Clusters}

In this section, we formalize our analysis of the clusters determined by the PCA and $k-nn$ assessment. We will discuss how the composition of independent campaign configuration variables contributes to different outcomes. Specifically, we will analyze whether an investment in the Fooji Delivery network provides meaningful differences in the performance of campaigns.

Our independent variables are set before a campaign begins and are not influenced by external factors (see Table \ref{table_independent_variables}). Meanwhile, dependent variables can be thought of as an outcome of the campaign. Our dependent variables are the number of \textit{Participations}, as a reference of the number of times a user-provided their personal information to receive a prize; and the \textit{Engagements}\footnote{\url{https://developer.twitter.com/en/docs/twitter-api/enterprise/engagement-api/overview}} the number of times a campaign was interacted with, defined as a like or comment, on Twitter. Participations is the primary objective of Fooji campaigns and we will use it as our operative metric. We believe these dependent variables are valued as a measure of a brand's ability to create stable relationships with its customers.

The table \ref{table_summary_clusters_v2} contains the mode of the categorical independent variables and mean values for numerical variables. Patterns in \texttt{fm\_categories} show a high dependence on this variable that generates cluster 1. Changes in the Delivery Networks for \texttt{giveaways\_categories} variable help to explain the division of Clusters 0 and 2. 
We believe the results show that progressive investment in Fooji's delivery network provides increased campaign outcomes.

\begin{table}[hbt!]
\centering
\begin{tabular}{lccc}
\hline
Independent variable  & $Cluster_0$ & $Cluster_1$  & $Cluster_2$ \\
\hline
\texttt{fm\_categories}$^{(c)}$        & Delivery            & Delivery + Mailed    & Delivery    \\
\texttt{giveaways\_categories}$^{(c)}$ & Delivery Network 1  & Delivery Network 1,  2 &  Delivery Network 3    \\
\texttt{pt\_categories}$^{(c)}$        & Delivery            & Delivery               & Delivery    \\
\texttt{sch\_categories}$^{(c)}$       & Twitter             & Twitter                & Twitter     \\
\texttt{cf\_total\_days}$^{(n)}$       & $3.00$              & $1.57$                 & $2.14$      \\
\texttt{ad\_total}$^{(n)}$             & $791.66$            & $1267.14$              & $379.33$    \\
\texttt{schp\_total}$^{(n)}$           & $3367.00$           & $3984.71$              & $1018.28$   \\
\hline
\end{tabular}
\caption{Summary of values in the clusters. The model was obtained for categorical variables $^{(c)}$ and the mean is the calculated value for numerical variables $^{(n)}$.  This table was generated using the 7 observations that are the closest to the centroid in every cluster.
 \label{table_summary_clusters_v2} }
\end{table}

 



Observing Cluster $0$, the mode of the campaigns in the cluster showed \textit{Delivery} as the Fulfillment option. However, the Giveaway Category was \textit{Delivery Network $1$}. This confirms the way that the clusters are divided using the second principal component at value $0$ in Y axis on Figure \ref{fig_cluster_analysis}. Cluster $0$'s combination of Delivery Fulfillment plus Delivery Network 1 means that it used the "Just-in-time" local delivery method to achieve its results. 
Cluster $0$ showed a mean campaign run time of $3.00$ days and a mean number of participations of $3367.00$.

In regards to Cluster $2$, the mode of the campaigns also showed \textit{Delivery}  as the Fulfillment option. However, the Giveaway Category was \textit{Delivery Network $3$} indicating that the campaign used an alternative delivery method. This alternative delivery method resulted in the worst performance of the groups. On average, these campaigns ran for $2.14$ days and produced a mean result of $1018.28$ participation.

Finally, Cluster $1$ shows the maximum returns of the Fooji Delivery network. These campaigns chose the \textit{Delivery} and \textit{Mailed} Fulfillment Methods. This indicates that they utilized Fooji's nationwide network configuration with automotive delivery in the local market and mailed delivery to participants outside the local market. \textit{Delivery Network 1} and \textit{Delivery Network 2} were the Giveaway categories selected.
On average, these campaigns ran for just $1.57$ days and show significantly a higher number of participation at $3984.71$.



\section{Conclusion}
 
    Our analysis applies machine learning techniques to create valid campaign clusters. Once these clusters are created, we analyzed their character to determine the underlying drivers of campaign engagement. The results of our analysis show that social media engagement is significantly improved by investment in "Just-in-time" delivery networks. 
    The difference in performance between maximum investment and minimum investment in the delivery network was 330.6 percent in favor of maximum investment. This confirms conventionally held opinions that; consumers prefer "Just-in-time" delivery and that "Just-in-time" delivery can create materially improve relationships with customers.
    
    The implications of these results extend beyond the Fooji social media business model. They support material investments in "Just-in-time" delivery based on objective measures of consumer satisfaction. In the future, we intend to measure the performance of specific goods in "Just-in-time" delivery to determine which consumer goods create the greatest benefits for corporations and consumers. In conclusion, the impact of on-demand Fulfillment methods and Giveaway methods on performance suggests that investment in real-time supply chain technologies has a strong potential to increase corporate sales, logistics productivity, and consumer satisfaction.

\bibliographystyle{unsrt}  
\bibliography{references}

\end{document}